\documentstyle[12pt,aaspp4]{article}

\lefthead{B\'ejar et al.}
\righthead{The Substellar Mass Function in $\sigma$ Orionis}
\slugcomment{To be published in ApJ Main Journal}

\newcommand{\mjup}{\,$M_{\rm Jup}$}
\newcommand{\msol}{\,$M_{\odot}$}
\newcommand{\sigori}{$\sigma$\,Orionis}

\begin{document}

\title{The Substellar Mass Function in $\sigma$ Orionis}

\author{V. J. S. B\'ejar}
\affil{Instituto de Astrof\'\i{}sica de Canarias, E-38205 La Laguna, Tenerife, 
Spain}

\author{E. L. Mart\'{\i}n}  
\affil{Institute of Astronomy. University of Hawaii at 
Manoa. 2680 Woodlawn Drive. Honolulu, HI 96822, USA}

\author{M. R. Zapatero Osorio }
\affil{Instituto de Astrof\'\i{}sica de Canarias, E-38205 La Laguna, Tenerife, 
Spain}
\affil{Current address: Division of Geological and Planetary Sciences, California Institute of 
Technology, Pasadena, USA}

\author{ R. Rebolo }
\affil{Instituto de Astrof\'\i{}sica de Canarias, E-38205 La Laguna, Tenerife, 
Spain}
\affil{Consejo Superior de Investigaciones Cient\'{\i}ficas, CSIC, Spain}

\author{D. Barrado y Navascu\'es}
\affil{Max-Planck-Institut f\"ur Astronomie, K\"onigstuhl 17, 
     D--69117 Heidelberg. Germany} 
\affil{Universidad Aut\'onoma de Madrid, E--28049 Madrid, Spain }

\author{C. A. L. Bailer-Jones, R. Mundt}
\affil{Max-Planck-Institut f\"ur Astronomie, K\"onigstuhl 17, 
     D--69117 Heidelberg, Germany} 

\author{I. Baraffe, C. Chabrier, F. Allard}
\affil{\'Ecole Normale Sup\'erieure, Lyon Cedex 7, France}

\centerline{e-mail addresses: vbejar@ll.iac.es, ege@teide.ifa.hawaii.edu, 
mosorio@gps.caltech.edu, rrl@ll.iac.es,} 
\centerline{barrado@pollux.ft.uam.es, calj@mpia-hd.mpg.de, 
mundt@mpia-hd.mpg.de,} 
\centerline{ibaraffe@ens-lyon.fr, chabrier@ens-lyon.fr, fallard@ens-lyon.fr}

\begin{abstract}

We combine results from imaging searches for substellar objects in the 
\sigori~ cluster  and follow-up photometric and spectroscopic observations    
to derive a census of the brown dwarf population in a region of 847 arcmin$^2$. 
We identify 64 very low-mass cluster member candidates in this region. 
We have available three color ($I$$Z$ and $J$) photometry for all of them, spectra for 9 objects, 
and $K$ photometry for 27 \%~of our sample. 
These data provide a well defined sequence in the $I$ versus $I-J$, $I-K$ color magnitude diagrams, 
and indicate that the cluster is affected by little reddening despite its young age ($\sim$5~Myr). 
Using state-of-the-art evolutionary models, we derive 
a mass function from the low-mass stars (0.2\msol) across the complete brown 
dwarf domain (0.075\msol~to 0.013\msol), and into the realm of free-floating planetary-mass objects 
($\le$ 0.013\msol). 
We find that the mass spectrum $(dN/dm) \propto m^{-\alpha}$ increases toward lower masses with an 
exponent $\alpha$ = 0.8 $\pm$ 0.4. 
Our results suggest that planetary-mass isolated objects could be as common as brown 
dwarfs; both kinds of objects together would be as numerous as 
stars in the cluster. If the distribution of stellar and substellar masses in 
\sigori~ is representative of the Galactic disk, 
older and much lower luminosity free-floating planetary-mass 
objects with masses down to about 0.005 \msol~should be abundant 
in the solar vicinity, with a density similar to M-type stars. 
\end{abstract}


\section{Introduction}

Although there is no definitive theory to explain the formation processes of 
stars, the widely accepted scenario 
is that they form via fragmentation of rotating interstellar molecular clouds 
followed by gravitational 
collapse. However, given the typical conditions 
and properties of Galactic molecular clouds, this simple paradigm has 
difficulties (Bodenheimer
\cite{bod98}) in explaining the genesis of numerous populations of substellar 
objects 
($M\,<\,0.075$\msol). Several arguments  have also been proposed against 
the formation of 
objects below the substellar boderline (Silk \cite{silk95}) or below the 
deuterium-burning mass limit 
(Shu, Adams \& Lizano \cite{shu87}), which according to Saumon 
et al. (\cite{saumon96}) and Burrows et al. (\cite{burrows97}) is located in the range 
0.013--0.011\msol~($\sim$ 14--12\mjup, where 1 \msol~= 1047 \mjup). 
The overall distribution of masses for invidual objects resulting from  star-forming 
processes can be described by the mass function (MF), defined as the the number 
of objects per interval of mass on  a 
logarithmic scale, $\xi(m) = dN/d\log m$,
 or alternatively by  the mass spectrum 
 defined as $\phi(m)$ = ${dN}/{dm}$. The MF was first studied for 
the stellar regime by Salpeter (\cite{salpeter55}), who found that a power-law 
relation of the type 
$\xi$(m) $\propto M^{- \gamma}$, 
with an index $\gamma$\,=\,1.35, (which corresponds to $dN/dm \propto m^{-\alpha}$ with 
$\alpha$ = 2.35 for the mass spectrum) was adequate in the mass range 1--10\msol. 
Subsequent studies of the field MF appear to demand lower values of $\alpha$ at smaller masses, 
or even suggest alternative functional forms (Miller \& Scalo \cite{miller79}). 
A recent study of the very low-mass MF based on DENIS and 2MASS discoveries of nearby 
ultracool dwarfs suggests a value 
of $\alpha$ in the range 1 to 2 (Reid et al. \cite{reid99}). 
A deep survey for methane dwarfs suggests, however, 
that $\alpha$ $\le$ 0.8 for disk brown dwarfs (Herbst et al. \cite{herbst99}).

Early searches for brown dwarfs in stellar clusters and associations (see eg. Rieke \& Rieke \cite{rieke90}; 
Stauffer, Hamilton \& Probst \cite{stauffer94}; Jameson \& Skillen \cite{jameson89}) and the subsequent confirmation of their existence (Rebolo, Zapatero Osorio, \& Mart\'{\i}n \cite{rebolo95}; Basri, Marcy \& Graham \cite{basri96}; Rebolo et al. 
\cite{rebolo96}) prompted among other questions the nature of the 
behavior of the MF in the brown dwarf domain and  
whether the fragmentation process can extend beyond the deuterium-burning mass
limit. Several studies in very young clusters have provided partial answers to these questions 
(Bouvier et al. \cite{bouvier98}; Luhman \& Rieke \cite{luhman99}; 
Luhman et al. \cite{luhman98}, \cite{luhman00}; Barrado y Navascu\'es et al. 
\cite{barrado01}; Tamura et al. 1998; Lucas \& Roche \cite{lucas00}; Hillenbrand \& Carpenter 
\cite{hille00}; Najita, Tiede \& 
Carr \cite{najita00}; Mart\'{\i}n et al. \cite{martin00}; Moreaux, Bouvier \& Stauffer \cite{moreaux01}). 
In spite of considerable progress made in recent years, the incompleteness of 
the photometric surveys at very low masses and the lack of a well 
defined spectroscopic sequence have prevented a 
reliable description of the MF over the whole brown dwarf regime. Here we present 
a determination of the MF for the 
\sigori~young stellar cluster, which is reliable and complete 
down to the deuterium-burning mass limit, and a first estimate on 
how this MF extends to smaller 
masses, i.e., to  the planetary regime. 

\section{ Age, distance, and extinction in the \sigori~cluster}

The \sigori~cluster belongs to the Orion OB1b association, for which an age of 
1.7--7\,Myr 
and a distance modulus of 7.8--8.3 are 
estimated based on studies carried out on massive stars (Blaauw 
\cite{blaauw64}, \cite{blaauw91}; Warren \& Hesser 
\cite{warren78}; Brown, de Geus \& de Zeeuw \cite{brown94}). The spectral type of the central star of 
the same name is O9.5 V. In order to account for the location of this star in the hydrogen-burning 
phase, its age must be younger than 5 Myr (on the basis of models with winds from Meynet et al. 
\cite{meynet94}). Recent investigations of the low-mass stellar and 
brown dwarf cluster populations have confirmed that \sigori~has indeed a very young age in 
the interval 1--5\, Myr 
(B\'ejar, Zapatero Osorio \& Rebolo \cite{bejar99} (BZOR); Wolk \& Walter \cite{wolk00}), 
which is consistent with the estimates found for the massive stars. 
The inferred MF in \sigori~may be very close 
to the true initial mass function (IMF) since no significant dynamical 
evolution is expected for cluster members. 
Additionally, the distance to the cluster is known through the determination 
provided by {\it Hipparcos\/}
 of the distance modulus of $m-M =  7.7 \pm 0.7$ (value given for the central star). This 
measurement is 
in agreement with previous distance determinations of the OB1b subgroup. 
The \sigori~star is affected 
by a low extinction of $E(B-V) = 0.05$ (Lee \cite{lee68}), thus, the 
associated cluster is expected to exhibit 
very little reddening. From the comparison of the colors of some of the 
\sigori~objects with counterparts of the same spectral type in the Pleiades 
and the field, BZOR did not find any significant reddening. In addition, the
location of a larger sample of objects in the  $I-J$ vs $J-K$ color-color diagram shows 
that their infrared excesss $E(I-J)$ is smaller than 0.3\,mag (i.e., $A_{V} \le 1$\,mag, on the basis 
of the relationships given in Rieke \& Lefobski \cite{rieke85}). 
All these properties of youth, proximity and low extinction confirm this cluster as 
a very interesting site for investigating the IMF.

\section{Surveys and membership selection criterion}

In order to construct the brown dwarf MF in the \sigori~ cluster we have 
combined optical ($IZ$) and near-infrared ($J$)  surveys
recently conducted around the central star (Zapatero Osorio et 
al. \cite{osorio00}; BZOR; B\'ejar \cite{bejar00}). 
New observations in the optical range were obtained with the Wide Field Camera 
instrument mounted on the primary focus of the 2.5--m Isaac Newton Telescope at the 
Roque de los Muchachos Observatory on 1998 November 12--13 (B\'ejar \cite{bejar00}). 
Images were bias-substracted and flat-fielded within the IRAF\footnote{IRAF is distributed 
by National Optical Astronomy Observatories, which is operated by the Association of Universities for 
Research in Astronomy, Inc., under contract with the National Science Foundation.} environment. 
Instrumental magnitudes 
were transformed into observed magnitudes by differential photometry of objects in common with images 
taken under photometric conditions with the IAC80 telescope (Teide Observatory), which were 
calibrated in the Cousins system by observing Landolt's (\cite{landolt92}) standard stars at different 
airmasses. 
Near infrared photometry in the $J$--band has been acquired using the 3.5--m 
telescope at the Calar Alto Observatory on 1998 October 27--31 (Zapatero Osorio et 
al. \cite{osorio00}). In addition, $K$--band photometry 
has been obtained on individual candidates with the 1.5--m Carlos S\'anchez Telescope at Teide Observatory (1998 
September 18, 2000 January 27, 2000 February 20), 
the 2.2--m telescope at Calar Alto Observatory (2000 February 16--18) 
and the 3.8--m United Kingdom Infrared telescope (UKIRT) at the Mauna 
Kea Observatory (2000 December 5--6). Raw frames were reduced following 
standard techniques in the infrared, which include sky-substraction and flat-fielding. 
The photometric calibration in the UKIRT system was achieved with faint standard stars 
(Hunt et al. \cite{hunt98}) observed at different airmasses on the same nights, 
except for the UKIRT data, which were calibrated later using objects in comon with images taken 
under photometric conditions with the 1.23--m telescope at Calar Alto Observatory during 2000 January 
22--23. 
The $I$$Z$ and $J$-band  
data of these surveys overlap in  a sky region of 847\,arcmin$^2$ (the location of this region is shown in 
Fig.1 of BZOR). 
Therefore we restrict our MF analysis to this particular region of the cluster in which 
we have three color photometry for all candidate members, with limiting $I_{\rm Cousins}$ and 
$J_{\rm UKIRT}$ magnitudes of 23.8 and 21.2, and completeness magnitudes of 21.5 and 
19.5, respectively. We have adopted as the limiting magnitude of our survey the
detection of 95\%~of the total number of point-like sources on the
frames; and as completeness magnitude the value at which the number
distribution of detections as a function of magnitude deviates from an
exponential law.  
 
Spectroscopic observations of a total of 14 candidates in 
\sigori~   have confirmed them as cluster members   (see BZOR, B\'ejar et al. \cite{bejar00}; 
Zapatero Osorio et al. \cite{osorio99}, \cite{osorio00}). We note that nine of them are located in 
the overlapping area of 847 arcmin$^2$. 
The 14 members give a well defined spectroscopic sequence 
from M6 (the most luminous and bluest targets) down to L4 (the reddest ones close to the 
limiting magnitude of the survey). 
Available $I$ and $J$-band observations for these objects
allow us to define the location of the low-mass star and brown dwarf 
sequence of the cluster (Figure~1), which we will adopt as a reference for the
identification of cluster members. This location is suitably reproduced
by the combination of the 5 Myr ``dust-free and dusty'' Lyon models 
(Baraffe et al. \cite{baraffe98}; Chabrier et al. \cite{chabrier00}) 
as shown in Figure~1. Below $I$\,=\,20 we expect dust condensates 
in the atmosphere of cluster members cooler than M9, and so the dusty models seem to be more appropiate.

In the 847 arcmin$^2$ region under consideration we identify a total of 64 photometric candidates 
that are distributed along the theoretical and observational sequences with a dispersion 
around 0.5\,mag. They seem to be very young objects and have colors redder than the 
10 Myr isochrone given by the dust-free Lyon models (see Fig.~1). All the candidates have $I-Z$ 
colors and $I$ magnitudes consistent with cluster 
membership. Follow-up $K$-band photometry for 17 of them also indicates their belonging to the 
$I$ vs $I-K$ cluster sequence, which reinforces their very likely membership 
(BZOR; B\'ejar et al. \cite{bejar00};  Zapatero Osorio et al. 
\cite{osorio00}). In addition, we have very recently obtained spectra for 6 of our faintest candidates; 
based on our preliminary analysis these objects fit the expected spectroscopic sequence 
and so are bona fide low-mass members with a very high probability (Barrado y Navascu\'es et al.
\cite{barrado01c}). 
The photometric and spectroscopic data of our candidates and those members defining the 
cluster sequence are shown in Tables~1 and ~2. In the latter we have not included the six candidates 
from Barrado y Navascu\'es et al. (\cite{barrado01c}). 
As explained in the previous section we do not find any evidence 
for reddening or infrared excesses and so we have not applied any extinction correction to our data. 
>From the successful  spectroscopic results along the photometric 
sequences  we conclude that our selection criterion using  optical and near-infrared photometry is 
very efficient in identifying true members  of the cluster.  A similar  criterion for membership
 has proved  successful in low-extinction clusters such as the
Pleiades (Zapatero Osorio  et al. \cite{osorio97}; Mart\'{\i}n et al. \cite{martin00}; 
Moreaux et al. \cite{moreaux01}) and IC 2391 (Barrado y Navascu\'es et al. \cite{barrado00}).

\section{The mass spectrum of brown dwarfs and planetary mass objects}

The cluster luminosity function (LF) have been derived by counting the number of objects per 
magnitude interval in the $I$ band, and it is shown in Figure~2. The first
bin, $M_{I}$ = 7.5--8.5, corresponds to stars so bright that were saturated
 in some of the images of the surveys under consideration. Fortunately, the BZOR
data allowed us to make an estimate of the counts for this massive bin which was conveniently normalized 
to the present survey. We can see in Figure~2 that the LF is rising up to $M_{I}$=9~mag 
and then falls down and becomes flat from $M_{I}$=11.5~mag. The LF remains flat down to 
the completeness limit of our surveys. 
We note that the bins where the luminosity function shows a peak correspond to a mass 
range (0.08--0.05 \msol) which includes objects that have  finished the deuterium burning
phase (the more massive ones) and those actually burning deuterium.
Both types of objects will have similar luminosities, if the  age of the cluster is in the range 
3--6 Myr, and therefore contribute to produce a peak in the LF. 

In order to derive the IMF, we have first determined the masses for the \sigori~members following 
a similar procedure to that 
described in Zapatero Osorio et al. (\cite{osorio00}), which means that we adopted the 
mass--luminosity relationship 
given by the Lyon models (Baraffe 
et al. \cite{baraffe98}; Chabrier et al. \cite{chabrier00}). In favor of these 
models it can be argued that 
they have been successful in fitting the mass--luminosity relation in various 
optical and infrared 
passbands (Baraffe et al. \cite{baraffe98}; Delfosse et al. \cite{delfosse01}), as well as in predicting coeval ages 
for the members of several young multiple systems (White et al. \cite{white99}; 
Luhman 1999), and that they provide a good fit to the infrared photometric sequence 
in the Pleiades and \sigori~clusters (Mart\'{\i}n et al. \cite{martin00}; 
Zapatero 
Osorio et al. \cite{osorio00}). Additionally, the Lyon tracks 
provide magnitudes and colors in the filters of interest as a function of mass, 
while in order 
to transform the effective temperatures and luminosities of other models into 
observables we would 
have to use bolometric corrections.

The \sigori~cluster substellar IMF is illustrated in Figure~3, where the mass 
spectrum representation on a logarithmic scale is 
provided. For the age of 5\,Myr a single power-law fit 
facilitates a reasonable 
representation of the data points with a slope of $\alpha$\,=\,0.8\,$\pm$\,0.4 
in the mass range 
which goes from very low mass stars (0.2\msol) through the whole brown dwarf 
domain to 0.013\msol. 
The uncertainty of $\pm$0.4 in the $\alpha$ index accounts for possible 
different ages of the 
cluster and the use of other evolutionary models. We have investigated the 
sensitivity of our 
mass spectrum to age by deriving $\alpha$ for ages from 3\,Myr to 10\,Myr. The 
values found were between 0.5 to 1.0.
This interval also accounts for an uncertainty of 0.2 
mag in the estimation of the cluster distance 
modulus. The dependence of the mass spectrum on theoretical models is even 
more 
uncertain. Our calculations considering Burrows et al. (\cite{burrows97}) 
isochrones yield  $\alpha$ values  up to 0.4 higher depending on age. 
Our main result is that the very low-mass stellar and substellar mass spectrum 
of the \sigori~ cluster is generally rising toward lower masses. 
IMFs with slopes in the range 0.4--0.8 below the star--substellar mass borderline,  have 
been obtained recently for other young clusters (Luhman et al. \cite{luhman00}; 
Lucas \& Roche \cite{lucas00}; Hillenbrand \& Carpenter \cite{hille00}; Najita 
et al. \cite{najita00}; 
Mart\'{\i}n et al. \cite{martin00}; Moreaux et al. \cite{moreaux01}), 
showing that the formation of brown dwarfs is a  quite common process in the Galactic disk. 

A remarkable feature of Figure~3 is the evidence for an extension of the IMF 
into the domain of planetary masses 
(i.e lower than the deuterium burning mass).
Despite the incompleteness of our survey and the possible 
contamination of 
several cluster non-members at these very low masses (see details in Zapatero 
Osorio et al. \cite{osorio00}), 
the planetary 
mass interval is rather well populated. This indicates that free-floating 
planetary mass objects 
with masses 0.013--0.005\msol~are abundant in \sigori. We find no evidence
 for a ``bottom end'' of the IMF 
in the mass interval covered by our analysis, i.e., there is no obvious deficit 
of objects near and beyond 
the deuterium-burning mass limit. Deeper surveys will be needed to determine the existence and location of a minimum-mass limit in the IMF. 

\section{Conclusions and future perspectives}

Recent searches have found a significant population of brown dwarfs in the 
$\sigma$ Orionis cluster. 
We have estimated the mass spectrum, $dN/dm \propto M^{-\alpha}$, from very low
mass stars (0.2\msol) to 0.013\msol~ and we have 
found that this is still rising across the whole brown dwarf regime with 
$\alpha$=0.8\,$\pm$\,0.4. Our results also suggest that the mass spectrum keeps rising down to 
0.005\msol . If the IMF in the \sigori~cluster has $\alpha$=0.8 down to 1 Jupiter mass, 
isolated planetary-mass objects in the mass range 1--12\mjup~ would be as 
numerous as brown dwarfs; and brown dwarfs and free-floating 
planets together 
would be as numerous as stars (see below for further details). 
However, their contribution to the total mass in the 
cluster would be less than 10~\%. 

The relatively large number of free-floating planetary-candidate members found 
in the \sigori~cluster 
suggests that such low-mass objects form commonly in nature, and that older and 
cooler isolated 
planets could be populating the Galactic disk and hence the solar neighborhood. 
Assuming that the IMF of \sigori~ 
is representative of the disk population, and extrapolating it to 
a mass of 1 \mjup, we obtain the densities of free-floating substellar systems 
given in Table~3. They are anchored to a density of stellar systems 
in the solar neighborhood of 0.057 $pc^{-3}$ (Reid et al. 1999). With this estimate for an index 
of $\alpha$ $\sim$ 1 in the mass spectrum we would expect a total 
number of substellar objects around 435 within a radius of 10 pc, whereas 
there would be 239 stars. Isolated planets much older than objects in \sigori~
will be extremely faint and cool enough to show molecular features like the 
giant planets in the Solar 
System. Therefore, even if they form a large population in the solar 
neighborhood, their detection 
is a challenge to present-day
 observational capabilities. According to theoretical 
predictions of radiated fluxes at different wavelengths (Burrows et al. 
\cite{burrows97}; Allard et al. 
\cite{allard97}), these objects in the mass range 1--12 \mjup~at the solar age could have 
effective temperatures of 100--300 K and an absolute magnitude of $M_{\rm J}$ = 20--25 and 
$M_{M}$=15--17. Current surveys such as 2MASS, DENIS, or SLOAN are unable to detect 
them out to distances greater than 1 parsec because they are too shallow. 
Deeper surveys, such as those reported by D'Antona, Oliva \& Zeppirelli (\cite{dantona99})  
and Herbst et al. (\cite{herbst99}) do not cover enough area. Free-floating planetary mass objects 
are extremely faint at optical and near-infrared wavelengths due to the 
absorption of methane and water, but they have a moderately transparent 
region around 5 $\mu$m. They could be identified with the {\it Space Infrared 
Facility\/} ({\it SIRTF\/}) out to distances of several parsecs from the Sun 
(Mart\'{\i}n et al. \cite{martin01}) or with wide ultra-deep ground-based near-infrared surveys such 
as the one planned with Megacam on the Canada-France-Hawaii telescope.

\acknowledgments

We are indebted to A. Burrows for facilitating an electronic 
version of his models. We are grateful to Carlos Guti\'errez and J. Licandro 
for taking data necessary for calibrating some of the $K$ images. Partial 
financial support was provided by the Spanish DGES project PB98--0531--C02--02. 
and CICYT grant ESP98--1339-CO2.

\clearpage


\figcaption[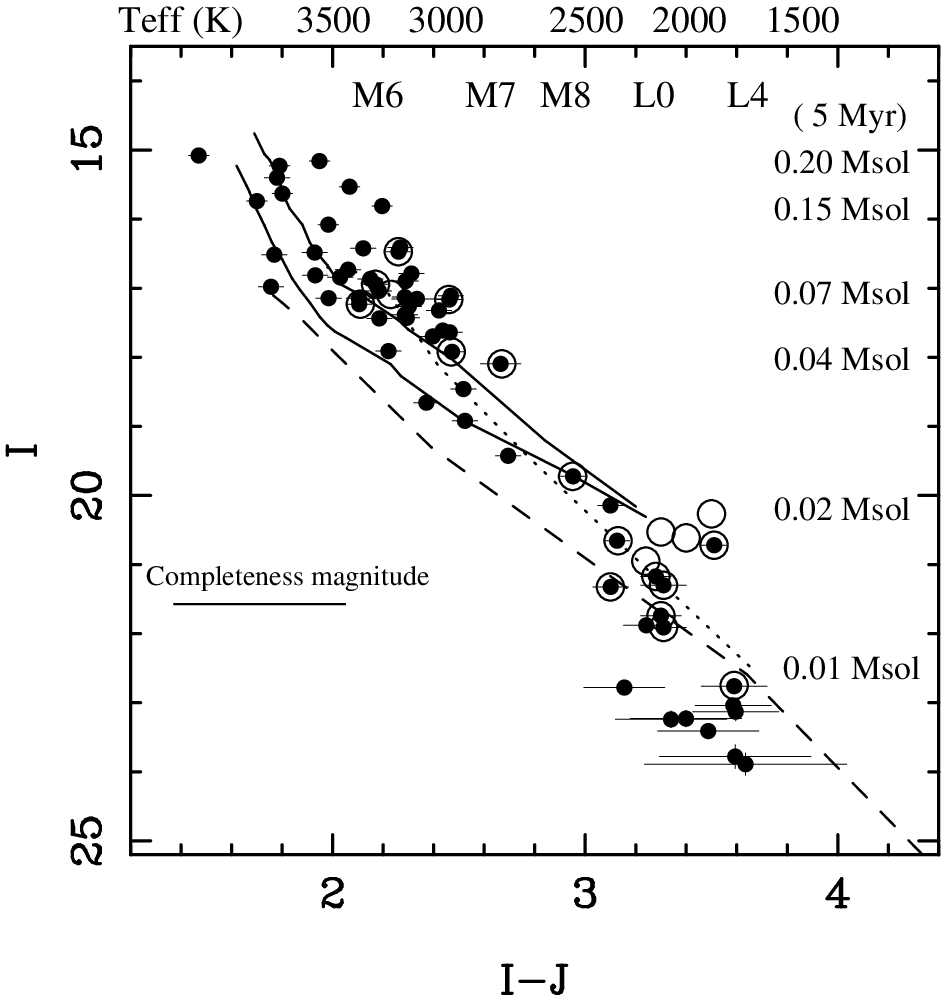]{\label{fig2} $I$ vs. $I-J$ color--magnitude 
diagram for the \sigori~cluster. Selected candidates are indicated 
with filled circles. The 5 Myr isochrones from the Lyon Group (Next Gen
 models---full line---and Dusty models---dashed line), 
and from the Arizona group (dotted line)  and the 10Myr Next Gen isochrone 
(full line, bluer than 5Myr) are also shown for 
comparison. Open circles around filled 
symbols denote candidates with available spectroscopy confirming their membership. 
Empty open circles are 
for members with spectroscopy but located outside of the 847 arcmin$^2$, and thus 
not included in the MF computation. Error-bars are based on photometric uncertainties 
and are smaller than symbol size for the majority of the brightest objects. 
Completeness magnitude, spectral type, estimated 
temperatures and masses for the age of 5 Myr are also shown.}

\figcaption[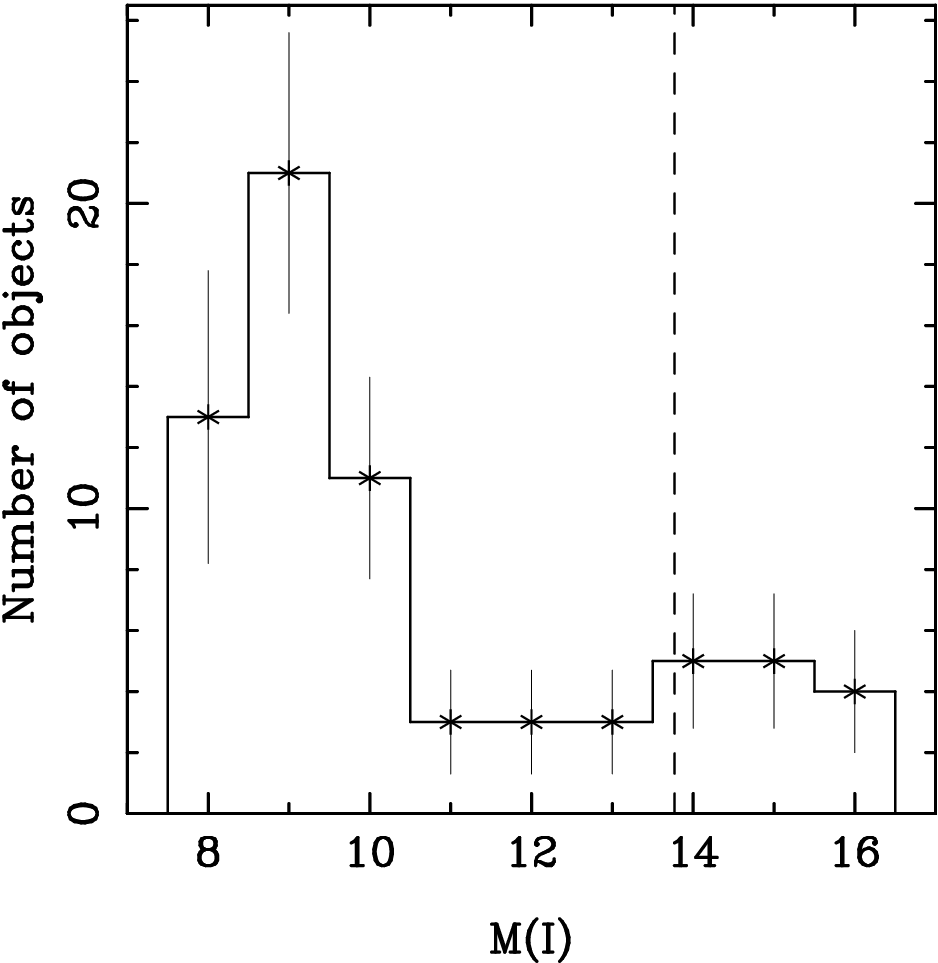]{\label{fig3} $I$-band luminosity function in the $\sigma$ 
Orionis cluster. the dashed line indicates the completeness limit of our search. 
Error bars corresponding to Poissonian uncertainties are also shown.}

\figcaption[newfig4.eps]{\label{fig4}
The mass function of the \sigori~cluster for substellar masses adopting 
several plausible ages. 
The best power-law fitting ($dN/dM \propto M^{-\alpha}$, dashed line) down to 
the brown dwarf-planet 
boundary ($\sim$ 0.013 \msol) gives $\alpha$\,=\,0.8\,$\pm$\,0.4 for the most probable age of 5 Myr. 
Error bars correspond to Poissonian 
uncertainties (from the finite number of objects), except for the planetary-mass 
interval where the upper limit (arrow) denotes the incompleteness of the photometric and 
spectroscopic searches, and 
the lower error bar accounts for the possible contamination of cluster 
non-members as discussed 
in Zapatero Osorio et al. (\cite{osorio00}).}

\clearpage
\plotone{fig2.eps}
\clearpage
\plotone{fig3.eps}
\clearpage
\epsscale{0.4}
\plotone{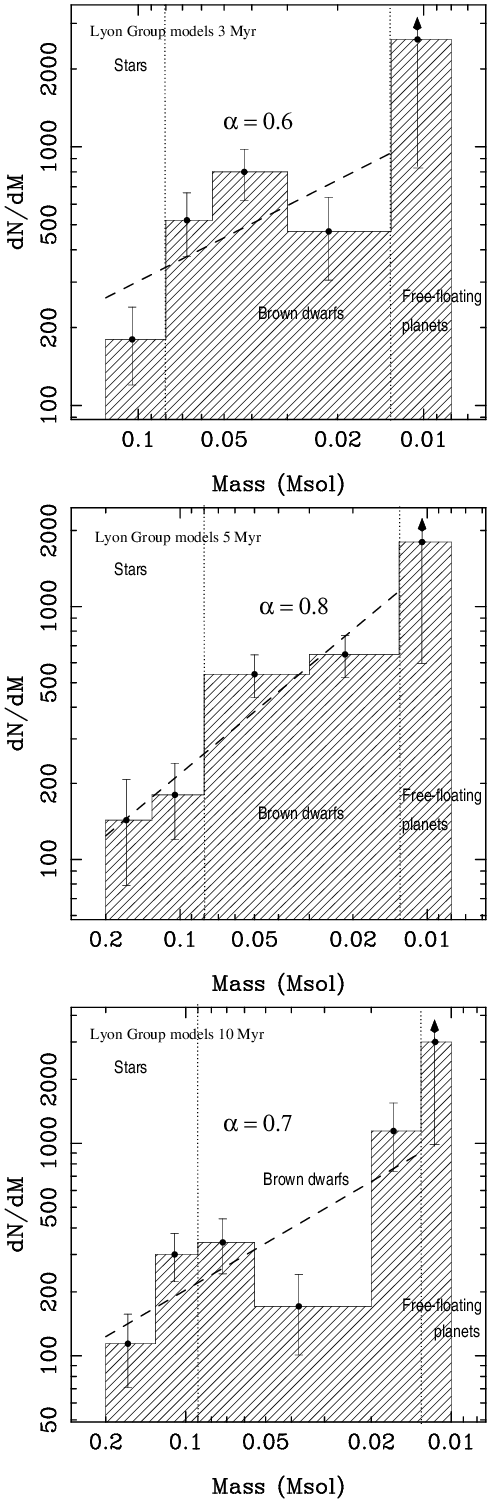}

\clearpage
\begin{deluxetable}{lccccccc}
\scriptsize
\tablecaption{\label{tab1} Photometric data of the selected candidates}
\tablewidth{0pt}
\tablehead{
\colhead{Name (IAU)} & \colhead{prev. ID.} & \colhead{I} 
& \colhead{R-I} & \colhead{I-J}  & \colhead{I-K}  & \colhead{R.A.(J2000)}  & \colhead{DEC.(J2000)}}
\startdata
 SOri J053911.7--022741 & SOri1  &  15.08$\pm$0.04   & 1.70$\pm$0.07  & 1.47$\pm$0.04   &                 & 05 39 11.7 &--02 27 41 \nl
 SOri J053920.8--023035 & SOri3  &  15.16$\pm$0.04   & 2.15$\pm$0.07  & 1.95$\pm$0.04   &                 & 05 39 20.8 &--02 30 35 \nl
 SOri J053939.2--023227 & SOri4  &  15.23$\pm$0.04   & 2.16$\pm$0.07  & 1.79$\pm$0.04   &                 & 05 39 39.2 &--02 32 27 \nl
 SOri J053920.1--023826 & SOri5  &  15.40$\pm$0.05   & 1.86$\pm$0.07  & 1.78$\pm$0.05   &                 & 05 39 20.1 &--02 38 26 \nl
 SOri J053847.5--023038 & SOri6  &  15.53$\pm$0.04   & 2.00$\pm$0.07  & 2.07$\pm$0.04   &                 & 05 38 47.5 &--02 30 38 \nl
 SOri J053908.1--023230 & SOri7  &  15.63$\pm$0.04   & 2.07$\pm$0.07  & 1.80$\pm$0.04   &                 & 05 39 08.1 &--02 32 30 \nl
 SOri J053907.9--022848 & SOri8  &  15.74$\pm$0.04   & 1.87$\pm$0.07  & 1.70$\pm$0.04   &                 & 05 39 07.9 &--02 28 48 \nl
 SOri J053817.1--022228 & SOri9  &  15.81$\pm$0.04   & 2.06$\pm$0.07  & 2.20$\pm$0.04   &                 & 05 38 17.1 &--02 22 28 \nl
 SOri J053944.4--022445 & SOri10 &  16.08$\pm$0.04   & 1.97$\pm$0.07  & 1.98$\pm$0.04   &                 & 05 39 44.4 &--02 24 45 \nl
 SOri J053944.3-023301  & SOri11 &  16.424$\pm$0.008 & 1.94$\pm$0.06  & 2.12$\pm$0.05   &                 & 05 39 44.3 &--02 33 01 \nl
 SOri J053757.4-023845  & SOri12 &  16.471$\pm$0.010 & 1.75$\pm$0.05  & 2.26$\pm$0.05   &                 & 05 37 57.4 &--02 38 45 \nl
 SOri J053813.1-022410  & SOri13 &  16.410$\pm$0.018 & 1.93$\pm$0.06  & 2.27$\pm$0.05   &                 & 05 38 13.1 &--02 24 10 \nl
 SOri J053909.9-022814  &        &  16.485$\pm$0.012 &                & 1.93$\pm$0.05   &                 & 05 39 09.9 &--02 28 14 \nl
 SOri J053746.6-024328  &        &  16.514$\pm$0.003 &                & 1.77$\pm$0.05   &                 & 05 37 46.6 &--02 43 28 \nl
 SOri J053911.4-023333  &        &  16.731$\pm$0.011 &                & 2.06$\pm$0.05   &                 & 05 39 11.4 &--02 33 33 \nl   
 SOri J053848.0-022854  & SOri15 &  16.789$\pm$0.014 & 1.81$\pm$0.07  & 2.31$\pm$0.05   & 3.31$\pm$0.06   & 05 38 48.0 &--02 28 54 \nl
 SOri J053849.2-022358  &        &  16.813$\pm$0.017 &                & 1.93$\pm$0.05   &                 & 05 38 49.2 &--02 23 58 \nl
 SOri J053915.0-024048  & SOri16 &  16.843$\pm$0.008 & 1.91$\pm$0.06  & 2.03$\pm$0.05   &                 & 05 39 15.0 &--02 40 48 \nl
 SOri J053721.0-022543  & SOri19 &  16.867$\pm$0.008 & 2.06$\pm$0.06  & 2.15$\pm$0.05   &                 & 05 37 21.0 &--02 25 43 \nl
 SOri J053825.6-023122  & SOri18 &  16.896$\pm$0.014 & 2.02$\pm$0.07  & 2.29$\pm$0.05   &                 & 05 38 25.6 &--02 31 22 \nl
 SOri J053904.4-023835  & SOri17 &  16.945$\pm$0.009 & 1.88$\pm$0.06  & 2.17$\pm$0.05   &                 & 05 39 04.4 &--02 38 35 \nl
 SOri J053923.3-024657  & SOri28 &  16.979$\pm$0.008 & 2.29$\pm$0.08  & 1.76$\pm$0.05   &                 & 05 39 23.3 &--02 46 57 \nl
 SOri J053829.0-024847  &        &  17.040$\pm$0.010 &                & 2.18$\pm$0.05   & 3.09 $\pm$0.03  & 05 38 29.0 &--02 48 47 \nl 
 SOri J053835.2-022524  & SOri22 &  17.109$\pm$0.008 & 2.11$\pm$0.07  & 2.47$\pm$0.05   &                 & 05 38 35.2 &--02 25 24 \nl
 SOri J053751.0-022610  & SOri23 &  17.128$\pm$0.009 & 2.10$\pm$0.06  & 2.29$\pm$0.05   &                 & 05 37 51.0 &--02 26 10 \nl
 SOri J053755.6-022434  & SOri24 &  17.144$\pm$0.009 & 2.01$\pm$0.06  & 2.10$\pm$0.05   &                 & 05 37 55.6 &--02 24 34 \nl
 SOri J053943.7-024729  & SOri32 &  17.144$\pm$0.007 & 2.26$\pm$0.07  & 1.98$\pm$0.05   &                 & 05 39 43.7 &--02 47 29 \nl
 SOri J053934.2-023847  & SOri21 &  17.154$\pm$0.007 & 1.91$\pm$0.08  & 2.33$\pm$0.10   &                 & 05 39 34.2 &--02 38 47 \nl
 SOri J053908.8-023958  & SOri25 &  17.163$\pm$0.008 & 2.17$\pm$0.10  & 2.46$\pm$0.05   &                 & 05 39 08.8 &--02 39 58 \nl
 SOri J053829.5-022517  & SOri29 &  17.230$\pm$0.008 & 1.98$\pm$0.07  & 2.11$\pm$0.05   &                 & 05 38 29.5 &--02 25 17 \nl
 SOri J053916.6-023827  & SOri26 &  17.264$\pm$0.008 & 1.83$\pm$0.08  & 2.30$\pm$0.05   &                 & 05 39 16.6 &--02 38 27 \nl
 SOri J053907.4-022908  & SOri20 &  17.321$\pm$0.009 & 1.68$\pm$0.07  & 2.42$\pm$0.05   &                 & 05 39 07.4 &--02 29 08 \nl
 SOri J053657.9-023522  & SOri33 &  17.385$\pm$0.008 & 2.28$\pm$0.06  & 2.29$\pm$0.05   &                 & 05 36 57.9 &--02 35 22 \nl
 SOri J053820.9-024613  & SOri31 &  17.429$\pm$0.008 & 2.03$\pm$0.05  & 2.29$\pm$0.05   &                 & 05 38 20.9 &--02 46 13 \nl
 SOri J053913.0-023751  & SOri30 &  17.438$\pm$0.008 & 1.71$\pm$0.08  & 2.19$\pm$0.05   &                 & 05 39 13.0 &--02 37 51 \nl
 SOri J053755.5-023308  & SOri35 &  17.612$\pm$0.008 & 2.25$\pm$0.06  & 2.44$\pm$0.05   &                 & 05 37 55.5 &--02 33 08 \nl
 SOri J053915.1-022152  & SOri38 &  17.640$\pm$0.008 & 2.19$\pm$0.09  & 2.46$\pm$0.05   &                 & 05 39 15.1 &--02 21 52 \nl
 SOri J053821.3-023336  &        &  17.697$\pm$0.013 &                & 2.40$\pm$0.05   &                 & 05 38 21.3 &--02 33 36 \nl
 SOri J053926.8-023656  & SOri36 &  17.911$\pm$0.008 & 1.94$\pm$0.14  & 2.22$\pm$0.05   &                 & 05 39 26.8 &--02 36 56 \nl
 SOri J053832.4-022958  & SOri39 &  17.922$\pm$0.008 & 2.24$\pm$0.10  & 2.47$\pm$0.05   &  3.18$\pm$0.07  & 05 38 32.4 &--02 29 58 \nl
 SOri J053736.4-024157  & SOri40 &  18.095$\pm$0.009 & 2.18$\pm$0.05  & 2.67$\pm$0.08   &                 & 05 37 36.4 &--02 41 57 \nl
 SOri J053936.4-023626  &        &  18.459$\pm$0.017 &                & 2.52$\pm$0.05   &                 & 05 39 36.4 &--02 36 26 \nl
 SOri J053926.8-022614  &        &  18.657$\pm$0.008 &                & 2.37$\pm$0.05   &                 & 05 39 26.8 &--02 26 14 \nl
 SOri J053948.1-022914  &        &  18.921$\pm$0.009 &                & 2.52$\pm$0.05   &                 & 05 39 48.1 &--02 29 14 \nl
 SOri J053912.8-022453  &        &  19.425$\pm$0.008 &                & 2.69$\pm$0.05   &  4.09 $\pm$0.10 & 05 39 12.8 &--02 24 53 \nl
 SOri J053825.6-024836  & SOri45 &  19.724$\pm$0.009 & 2.75$\pm$0.017 & 2.95$\pm$0.05   &  4.07 $\pm$0.09 & 05 38 25.6 &--02 48 36 \nl
 SOri J053946.5-022423  &        &  20.144$\pm$0.010 &                & 3.10$\pm$0.05   &  4.21 $\pm$0.16 & 05 39 46.5 &--02 24 23 \nl 
 SOri J053910.8-023715  & SOri50 &  20.656$\pm$0.015 &                & 3.13$\pm$0.05   &  4.48 $\pm$0.05 & 05 39 10.8 &--02 37 15 \nl
 SOri J053903.2-023020  & SOri51 &  20.72 $\pm$0.014 &                & 3.51$\pm$0.05   &  4.58 $\pm$0.10 & 05 39 03.2 &--02 30 20 \nl 
 SOri J053825.1-024802  & SOri53 &  21.172$\pm$0.023 &                & 3.28$\pm$0.06   &  4.72 $\pm$0.09 & 05 38 25.1 &--02 48 02 \nl
 SOri J053833.3-022100  & SOri54 &  21.30 $\pm$0.05  &                & 3.31$\pm$0.09   &  4.35 $\pm$0.10 & 05 38 33.3 &--02 21 00 \nl
 SOri J053725.9-023432  & SOri55 &  21.32 $\pm$0.03  &                & 3.10$\pm$0.07   &  4.32 $\pm$0.10 & 05 37 25.9 &--02 34 32 \nl
 SOri J053900.9-022142  & SOri56 &  21.74 $\pm$0.03  &                & 3.30$\pm$0.08   &  4.65 $\pm$0.10 & 05 39 00.9 &--02 21 42 \nl
 SOri J053947.0-022525  & SOri57 &  21.88 $\pm$0.03  &                & 3.24$\pm$0.09   &                 & 05 39 47.0 &--02 25 25 \nl
 SOri J053903.6-022536  & SOri58 &  21.91 $\pm$0.03  &                & 3.31$\pm$0.09   &  5.03 $\pm$0.20 & 05 39 03.6 &--02 25 36 \nl
 SOri J053937.5-023042  & SOri60 &  22.76 $\pm$0.05  &                & 3.59$\pm$0.13   &  5.07 $\pm$0.10 & 05 39 37.5 &--02 30 42 \nl
 SOri J053852.6-022846  & SOri61 &  22.78 $\pm$0.05  &                & 3.16$\pm$0.16   &                 & 05 38 52.6 &--02 30 46 \nl
 SOri J053942.1-023031  & SOri62 &  23.04 $\pm$0.07  &                & 3.59$\pm$0.15   &  5.36 $\pm$0.15 & 05 39 42.1 &--02 30 31 \nl
 SOri J053653.3-022414  & SOri64 &  23.13 $\pm$0.13  &                & 3.60$\pm$0.17   &  4.51 $\pm$0.25 & 05 36 53.3 &--02 24 14 \nl
 SOri J053724.7-023152  & SOri66 &  23.23 $\pm$0.12  &                & 3.40$\pm$0.22   &                 & 05 37 24.7 &--02 31 52 \nl
 SOri J053826.1-022305  & SOri65 &  23.24 $\pm$0.12  &                & 3.34$\pm$0.22   &  4.41 $\pm$0.30 & 05 38 26.1 &--02 23 05 \nl
 SOri J053812.6-022138  & SOri67 &  23.41 $\pm$0.090 &                & 3.49$\pm$0.20   &                 & 05 38 12.6 &--02 21 38 \nl
 SOri J053839.1-022805  & SOri68 &  23.78 $\pm$0.17  &                & 3.6 $\pm$0.3    &                 & 05 38 39.1 &--02 28 05 \nl
 SOri J053918.1-022855  & SOri69 &  23.89 $\pm$0.16  &                & 3.6 $\pm$0.4    &                 & 05 39 18.1 &--02 28 55 \nl

\enddata

\tablecomments{ Units of right ascension (J2000) are hours, minutes, 
and seconds, and units of declination (J2000) are degrees, arcminutes, 
and arcseconds. Coordinates are accurate to $\pm$1\arcsec. 
All the available $R$-band photometry and $I$-band data for candidates SOri1--10 have been taken 
from BZOR. Photometric meausurements for candidates SOri50-69 have also been presented in Zapatero 
Osorio et al. 2000.}
\end{deluxetable}

\clearpage

\begin{deluxetable}{lccccc}
\tablecaption{\label{tab2} Spectroscopic data of \sigori~ members}
\tablewidth{0pt}
\tablehead{ 
\colhead{Name} & \colhead{I} &  \colhead{I-J} & \colhead{I-K} & \colhead{Spectral} & \colhead{Spectral}\nl
\colhead{}     & \colhead{}  &  \colhead{}    & \colhead{}    & \colhead{Type (PC3)}         & \colhead{Type}             
}
\startdata
SOri12\tablenotemark{*}                & 16.471$\pm$0.010 & 2.26 $\pm$0.05  &                 & M4.5 & M6 \nl 
SOri17\tablenotemark{*}                & 16.945$\pm$0.009 & 2.17 $\pm$0.05  &                 & M4.6 & M6  \nl 
SOri29\tablenotemark{*}                & 17.230$\pm$0.008 & 2.11 $\pm$0.05  &                 & M4.8 & M6  \nl 
SOri25\tablenotemark{*}                & 17.163$\pm$0.008 & 2.46 $\pm$0.05  &                 & M5.1 & M6.5 \nl 
SOri39\tablenotemark{*}                & 17.922$\pm$0.008 & 2.47 $\pm$0.08  & 3.18$\pm$0.07   & M5.1 & M6.5 \nl 
SOri27                 & 17.090$\pm$0.04  & 2.23 $\pm$0.05  & 3.18$\pm$0.05   & M5.1 & M7 \nl 
SOri40\tablenotemark{*}                & 18.095$\pm$0.009 & 2.67 $\pm$0.06  &                 & M5.6 & M7 \nl 
SOri45\tablenotemark{*}                & 19.724$\pm$0.009 & 2.95 $\pm$0.05  & 4.07$\pm$0.09   & M8.0 & M8.5 \nl 
SOriJ053710.0-024302   & 20.266$\pm$0.011 & 3.5  $\pm$0.3   & 4.9 $\pm$0.4    & M8.2 & M8.5 \nl 
SOriJ053636.3-024626   & 20.614$\pm$0.019 & 3.4  $\pm$0.11  &                 & M9.4 & M9.5\nl
SOri47                 & 20.530$\pm$0.05  & 3.30 $\pm$0.10  & 4.79$\pm$0.15   & L1.4 & L1.5 \nl
SOri52                 & 20.958$\pm$0.016 & 3.24 $\pm$0.15  & 5.53$\pm$0.15   & L0.5 & L0.5 \nl
SOri56\tablenotemark{*}                & 21.740$\pm$0.03  & 3.30 $\pm$0.08  & 4.65 $\pm$0.10  & L0.5 & L0.5 \nl
SOri60\tablenotemark{*}                & 22.76 $\pm$0.05  & 3.59 $\pm$0.13  & 5.07 $\pm$0.10  &      & L4   \nl 

\enddata
\tablecomments{ Spectral type have been derived using pseudocontinuous index PC3 
([823.0--827.0]/[754.0--758.0], Mart\'{\i}n et al. 
1999) and from comparison with standard M dwarfs.} 
\tablenotetext{*}{ Candidates within the 847 arcmin$^2$ of present survey.}
\end{deluxetable}

\clearpage

\begin{deluxetable}{cccccc}
\tablecaption{\label{tab3} Substellar density in the solar vicinity}
\tablewidth{0pt}
\tablehead{
\colhead{$\alpha$}  & \colhead{$\rho_{\rm BD}$}  &  \colhead{$N_{\rm BD}$}  
&\colhead{$\rho_{\rm Pl}$}  &  \colhead{$N_{\rm Pl}$} 
&  \colhead{$N_{\rm tot}$} \nl
 &\colhead{systems/pc$^3$}  & \colhead{$d <$ 10 pc} & \colhead{systems/pc$^3$}  
& \colhead{$d <$ 10 pc} 
 & \colhead{$d <$ 10 pc} } 
\startdata
0.5  & 0.015 &  63 &  0.008 &   34 &   95\nl
0.8  & 0.028 & 117 &  0.027 &  113 &  230\nl
1.0  & 0.042 & 176 &  0.062 &  259 &  435\nl
1.5  & 0.114 & 478 &  0.510 & 2136 & 2614\nl

\enddata
\tablecomments{ $\alpha$ indicates the exponent of the mass spectrum ($dN/dm \propto m^{-\alpha}$) and 
BD and Pl indicates brown dwarfs (0.075--0.013 \msol) and planetary mass objects (0.013--0.001 \msol), respectively}
\end{deluxetable}

\end{document}